\pdfoutput=1
\documentclass[letterpaper,twocolumn,10pt]{article}
\usepackage{usenix}

\usepackage{tikz}
\usepackage{amsmath}
\usepackage{tabulary}
\usepackage{tikz,pgfplots,pgfplotstable,tikzscale,siunitx, xcolor}
\usetikzlibrary{patterns}
\usepackage{multicol}
\usepackage{filecontents}
\usepackage{graphicx}
\usepackage{flushend}

\usepackage{enumitem}
\newcommand{\ignore}[1]{}
\usepackage[pass]{geometry}
\usepackage[normalem]{ulem}
\usepackage{color}
\usepackage{soul}
\usepackage{subcaption}

\usepackage[algo2e,ruled,vlined]{algorithm2e}

\begin{document}
\title{\Large \bf Hardware Memory Management for Future Mobile Hybrid Memory Systems}
\author{
{\rm Fei Wen}\\
{\rm feitamu@gmail.com}
\and
{\rm Mian Qin}\\
{\rm celery1124@tamu.edu}
\and
{\rm Paul Gratz}\\
{\rm pgratz@gratz1.com}
\and
{\rm Narasimha Reddy}\\
{\rm reddy@tamu.edu}\\
\and
Department of Electrical \& Computer Engineering\\
Texas A\&M University
} 

\maketitle
\begin{abstract}
  The current mobile applications have rapidly growing memory
  footprints, posing a great challenge for memory system design.
  Insufficient DRAM main memory will incur frequent data swaps between
  memory and storage, a process that hurts performance, consumes
  energy and deteriorates the write endurance of typical flash storage
  devices.  Alternately, a larger DRAM has higher leakage power and
  drains the battery faster.  Further, DRAM scaling trends make
  further growth of DRAM in the mobile space prohibitive due to cost.
  Emerging non-volatile memory (NVM) has the potential to alleviate
  these issues due to its higher capacity per cost than DRAM and
  minimal static power.  Recently, a wide spectrum of NVM
  technologies, including phase-change memories (PCM), memristor, and
  3D XPoint have emerged.  Despite the mentioned advantages, NVM has
  longer access latency compared to DRAM and NVM writes can incur
  higher latencies and wear costs.  Therefore integration of these new
  memory technologies in the memory hierarchy requires a fundamental
  rearchitecting of traditional system designs.  In this work, we
  propose a hardware-accelerated memory manager (HMMU) that addresses
  both types of memory in a flat space address space.  We design a set
  of data placement and data migration policies within this memory
  manager, such that we may exploit the advantages of each memory
  technology.  By augmenting the system with this HMMU, we reduce the
  overall memory latency while also reducing energy consumption and writes to the NVM.
  Experimental results show that our design achieves a 39\% reduction
  in energy consumption with only a 12\% performance degradation
  versus an all-DRAM baseline that is likely untenable in the future.
\end{abstract}
\section{Introduction}
\label{sec:intro}

As the demand for mobile computing power scales, mobile applications
with ever-larger memory footprints are being developed, such as
high-resolution video decoding, high-profile games, etc.  This trend
creates a great challenge for current memory and storage system design.  The historical approach to address memory
footprints larger than the DRAM available is for the OS to swap less
used pages to storage, keeping higher locality pages in memory.  Given
the latencies of modern storage systems (even "high" performance
SSDs~\cite{intel-nvmessd, Strata, FlashShare}) are several orders of
magnitude higher than DRAM.However, allowing any virtual memory
swapping to storage implies incurring a severe slowdown.  Thus mobile
device manufacturer rapidly expanded the DRAM size for the worst case
possible memory footprint. For example, the DRAM capacity of the flagship phones from the Samsung Galaxy S series have expanded by 16X over the past ten years.  While this approach has been largely successful
to date, the size of DRAM is constrained by both cost/economics and
energy consumption.  Unlike data centers, mobile devices are highly
cost-sensitive and have a highly limited energy budget.  Moreover, the
DRAM technology has a substantial background power, constantly
consuming energy even in idle due to its periodic refresh requirement,
which scales with DRAM capacity.  Therefore a larger DRAM means a
higher power budget and a shorter battery life, particularly given
recent hard DRAM VLSI scaling limits.  The approach of provisioning
more DRAM is not sustainable and hard limits will soon be hit on the
scaling of the future mobile memory system.

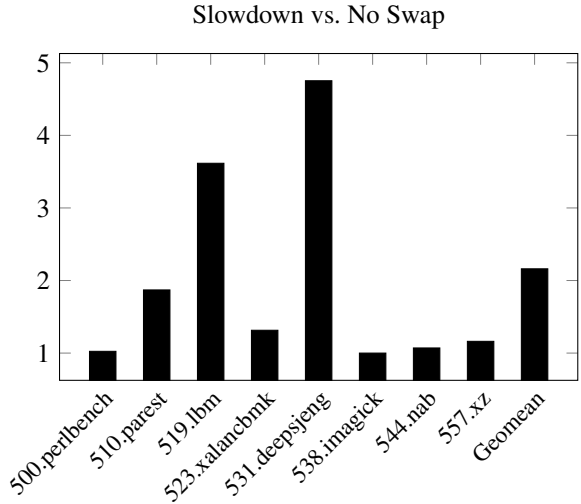
\begin{figure}[h]
  \centering
\begin{tikzpicture}
    \begin{axis}[
        width=\columnwidth, height=0.7\columnwidth,
        title={Slowdown vs. No Swap},
        symbolic x coords=
        {500.perlbench,510.parest,519.lbm,523.xalancbmk,531.deepsjeng,538.imagick,544.nab,557.xz,Geomean},
        xtick=data,
        xticklabel style={font=\small, rotate=45, inner sep=2pt, anchor=north east},
        ytick={0,1,2,3,4,5},
      ]
        \addplot[ybar,fill=black] coordinates {
        (500.perlbench,1.02417962)(510.parest,1.87019555)(519.lbm,3.615523466)(523.xalancbmk,1.314491264)(531.deepsjeng,4.75265215)(538.imagick, 1)(544.nab,1.070671378)(557.xz,1.161843516)(Geomean,2.161195403)
        };
    \end{axis}
\end{tikzpicture}
\caption{Performance Impact of OS Memory Management.}
\label{fig:OS_overhead}
\end{figure}

The emergence of several Non-Volatile-Memory (NVM) technologies, such
as Intel 3D Xpoint~\cite{intel-3dxpoint}, memristor~\cite{memristor},
Phase-change-memory(PCM)~\cite{PCM}, provides a new avenue to address
this growing problem.  These new memory devices promise an order of
magnitude higher density~\cite{techinsights} per cost and lower static
power consumption than traditional DRAM technologies, however, their
access delay is significantly higher, typically also within one order
of magnitude of DRAM.  Further, these new technologies show
significant overheads associated with writes and are non-volatile.
Thus, these emerging memory technologies present a unique opportunity
to address the problems of growing application workload footprints
with hybrid memory systems composed of both DRAM and emerging NVM
memories.  To exploit these new memory devices effectively, however,
we must carefully consider their performance characteristics relative
to existing points in the memory hierarchy.  In particular, while
memory access and movement in prior storage technologies such as flash
and magnetic disk is slow enough that software management via the OS
was feasible.  With emerging NVM memory accesses at within an order of
magnitude of DRAM, relying on traditional OS memory management
techniques for managing placement between DRAM and NVM is insufficient
as illustrated in Figure~\ref{fig:OS_overhead}.

In Figure~\ref{fig:OS_overhead}, a subset of benchmarks from the SPEC
CPU2017 benchmark suite are executed in a system where around 128MB of
the application's memory footprint is able to fit in the system DRAM
directly.  A ramdisk based swap file is set up to hold the remainder
of the application memory footprint.  Since this ramdisk swapfile is
implemented in DRAM it represents an upper bound on the performance
for pure software swapping.  The results shown are normalized against
a system where sufficient DRAM is available to capture the entire
memory footprint.  As we see, in this arrangement, the cost of pure OS
managed swapping to NVM would be quite high, with applications seeing
an average of $\sim$2X slowdown versus baseline.  As we will show, a
significant fraction of this overhead comes explicitly from the costs
of the required page fault handling.

Some existing work has begun to explore system design for emerging
hybrid memories.  Broadly this prior work falls into one of two
categories, first, some advocate using DRAM as a pure hardware managed
cache for NVM~\cite{Qureshi:2009,CAMEO}.  This approach implies a high
hardware cost for metadata management and imposes significant capacity
and bandwidth constraints.  Second, some have advocated for a purely
software, OS managed approach~\cite{Hassan:2015, span, Wang:2014}.  As
we discussed previously, this approach implies significant slowdowns
due to software overhead of the operating system calls.

Here we propose a new, hardware managed hybrid memory management
scheme which retains the performance benefits of caching, without the
high metadata overhead such an approach implies.  Compared to previous
work, our project has the following advantages:
\begin{itemize}[noitemsep,nolistsep]
\item With a ratio of 1/8 DRAM vs 7/8 NVM, we achieved 88\% of the
  performance of an untenable full DRAM configuration, while reducing
  the energy consumption by 39\%.
\item Compared to inclusive DRAM caches, we preserve the full main
  memory capacity for the user applications.
\item Parallel access to both the DRAM and NVM is supported, rendering
  a higher effective memory bandwidth. This also helps to suppress the
  excessive cache insertion/replacements and prevent cache thrashing.
\item The data placement and migration are executed by hardware.  This
  eliminates the long latency incurred by the OS managed virtual
  memory swap process.
\item Memory management and allocation are performed with a
  combination of page and sub-page-block sizes to ensure the best
  utilization of the available DRAM and to reduce the number and
  impact of writes to the NVM.
\end{itemize}
\begin{table*}[h]
  \centering
  \small
\caption{Approximate Performance Comparison of Different Memory
  Technologies\cite{NVM1,NVM2,yang:2012} }
\begin{tabular}{c|c|c|c|c|c|c}
\hline
Technology& HDD & FLASH & 3D XPoint & DRAM & STT-RAM & MRAM\\
\hline
Read Latency & 5ms & $100\mu s$  & 50 - 150ns & 50ns & 20ns & 20ns\\
Write Latency & 5ms & $100\mu s$  & 50 - 500ns & 50ns & 20ns & 20ns\\
Endurance (Cycles) & $>10^{15}$ & $10^{4}$ & $10^9$ & $>10^{16}$ & $>10^{16}$ & $>10^{15}$\\
\$ per GB & 0.025-0.5 & 0.25-0.83 & 6.5~\cite{nvm_price} & 5.3-8 & N/A & N/A\\
Cell Size & N/A & $4-6F^2$ & $4.5F^2$ ~\cite{techinsights} & $10F^2$ & $6-20F^2$ & $25F^2$\\
\hline 
\end{tabular}
\label{tab:nvmchar}
\end{table*}

\section{Background and Motivation}
\label{sec:motiv}
With emerging non-volatile memory technologies providing more memory
system capacity, density, and lower static power, they have the
potential to meet the continuously increasing memory usage of mobile applications.
Given their different characteristics from traditional DRAM and
storage, however, the design of systems comprising these new
technologies together with traditional DRAM and storage is an open
question.  Here we examine the characteristics of these new memory
technologies and the existing proposals to date on how to leverage
them in system designs.
\subsection{Nonvolatile Memory Technology Characteristics}
\label{sec:nvmtech}
Table~\ref{tab:nvmchar} shows the relative characteristics of several
emerging non-volatile memory technologies against traditional DRAM and
storage~\cite{ITRS2015,NVM1,NVM2}.  While HDD and
Flash have 100k and 2k times larger read access latency than DRAM
respectively, the emerging NVM technologies have read
access latencies typically within one order of magnitude of DRAM.
Meanwhile emerging non-volatile memory technologies provide higher memory
system capacity, density and lower static power. 

Given their different characteristics from traditional DRAM and
storage, however, the design of systems comprising these new
technologies together with traditional DRAM and storage is an open
question.  Here we examine the characteristics of these new memory
technologies and the existing proposals to date on how to leverage
them in system designs.

Further, we note that in these new technologies writes are often more
expensive that reads both in terms of latency as shown and
endurance/lifetime cost, as well as energy consumption for writing.

The relative closeness in performance and capacity to traditional DRAM
of emerging NVM technologies argues for a different approach to memory
management than traditional, OS or hardware-cache based approaches.
In the remainder of this section, we examine the prior work approaches
to the design of hybrid memory systems.

\subsection{Operating System-Based Memory Management}
Hassan \emph{et al.}, Fedorov \emph{et al.} and propose to leverage
the OS to manage placement and movement between NVM and
DRAM~\cite{Hassan:2015,span}.  They treat NVM as a parallel memory
device on the same level as that of DRAM in the memory hierarchy.
They argue that this approach can yield better utilization of the
large NVM capacity without wasting the also relatively large DRAM
capacity.  Their approach is similar to the traditional approach of
using storage as a swap space to extend the DRAM main memory space.
Direct application of this approach to NVM creates some difficulties,
however.  When a given requested data is found to be in the swap space
on the NVM, a page fault occurs which must be handled by operating
system.  The latency of this action is not only comprised of the
device latency itself but also the induced OS context switch, and page fault handling.  While in traditional storage systems with
\emph{ms}-level latencies, that cost is negligible, with the latency
of SSD and other NVM devices significantly decreased, the OS management overheads come to
dominate this latency, as discussed previously and indicated in Figure~\ref{fig:OS_overhead}.

\subsection{Hardware-managed DRAM Caches and Related Approaches}
\label{sec:dramcache}
Other groups have proposed using DRAM as the cache/buffer for NVM, and
thus turning DRAM into the new last level cache\cite{Qureshi:2009}.
Similar schemes have also been applied to other memory devices with
latency discrepancy in heterogeneous-memory-system(HMS). For instance,
3D-stacked DRAM was proposed as a cache for off-chip DRAM in the
works\cite{DRAM-cache1,Unison-cache,Alloy-cache,J.Sim}.  A common
theme in all these designs is the difficulty in lookup and maintenance
of the tag storage, since the number of tags scales linearly with the
cache size.  
Assuming the cache block size is 64B and 8 bytes of tag
for each block, then a 16GB DRAM cache requires 2GB for the tag
storage alone.  That is much too large to fit in a fast, SRAM tag
store.  Much of the prior work explores mechanisms to shrink the tag
storage overhead~\cite{Meza}.  Some researchers explored tag
reduction~\cite{tag-reduction}.  Others aimed to reconstruct the cache
data structure. For instance, some works combine the tag or other
meta-data bits into the data entry
itself~\cite{Unison-cache,Dram-cache2}.

Another issue these works attempt to address is the extended latency
of tag access.  DRAM devices have significantly greater access latency
than SRAM.  Additionally, their larger cache capacity requires a
longer time for the tag comparison and data selection hardware.  If
the requested data address misses in the TLB, it takes two accesses to
the DRAM before the data can be fetched. Lee \emph{et al.} attempted
to avoid the tag comparison stage entirely by setting the cache block
size to equal the page size, and converting virtual addresses to cache
addresses directly in a modified TLB~\cite{Tagless}.  This approach,
however requires several major changes to the existing system
architecture including requiring extra information bits in the page
table, modifying the TLB hardware and an additional global inverted
page table.

Broadly, several issues exist with the previously proposed,
hardware-based management techniques for future hybrid memory systems.

\begin{itemize}[noitemsep,nolistsep]
\item As with traditional processor cache hierarchies, every memory
  request must go through the DRAM cache before accessing the NVM.
  Prior work shows that this approach is sub-optimal for systems where
  bandwidth is a constraint and where a parallel access path is
  available to both levels of memory~\cite{Wu:2009}.  Further, given
  the relatively slow DRAM access latency requiring a miss in the DRAM
  before accessing the NVM implies a significantly higher overall
  system latency.
\item These works largely assume an inclusive style caching.  Given
  the relative similarity in capacity between DRAM and NVM, this
  implies a significant loss of capacity.
\item Given the capacities of DRAM and NVM versus SRAM used in
  processor caches, a traditional cache style arrangement implies a
  huge overhead in terms of cache meta-data.  This overhead will add
  significant delays to the critical path of index search and tag
  comparison, impacting every data access.
\end{itemize}

Liu \emph{et al.} propose a hardware/software, collaborative approach
to address the overheads of pure software approaches without some of
the drawbacks of pure hardware caching~\cite{Liu:2017}.  Their
approach, however, requires modifications both to the processor
architecture as well as the operating system kernel.  These
modifications have a high NRE cost and hence is difficult to be
carried out in production.

In this paper, we propose a hardware-based hybrid memory controller
that is transparent to the user and as well as the operating system, thus
it does not incur the overheads of management of OS based approaches.
The controller is an independent module and compatible with existing
hardware architectures and OSes.  The controller manages both DRAM and
NVM memories in flat address space to leverage the full capacity of
both memory classes.  Our approach also reserves a small portion of
the available DRAM space to use as a hardware-managed cache to
leverage spacial locality patterns seen in real application workloads
to reduce writes to the NVM.
\section{Design}
\label{sec:design}

Here we describe the proposed design of our proposed hardware memory
management for future hybrid memory systems.  Based on the discussion
in Section~\ref{sec:motiv} and cognizant of the characteristics of
emerging NVM technologies, we aim to design a system in which the
latency overheads of OS memory management are avoided, while hardware
tag and meta-data overheads of traditional caching schemes are minimized.

\subsection{System Architecture Overview}
\begin{figure}[!hbt]
\centerline{\includegraphics[width=\columnwidth]{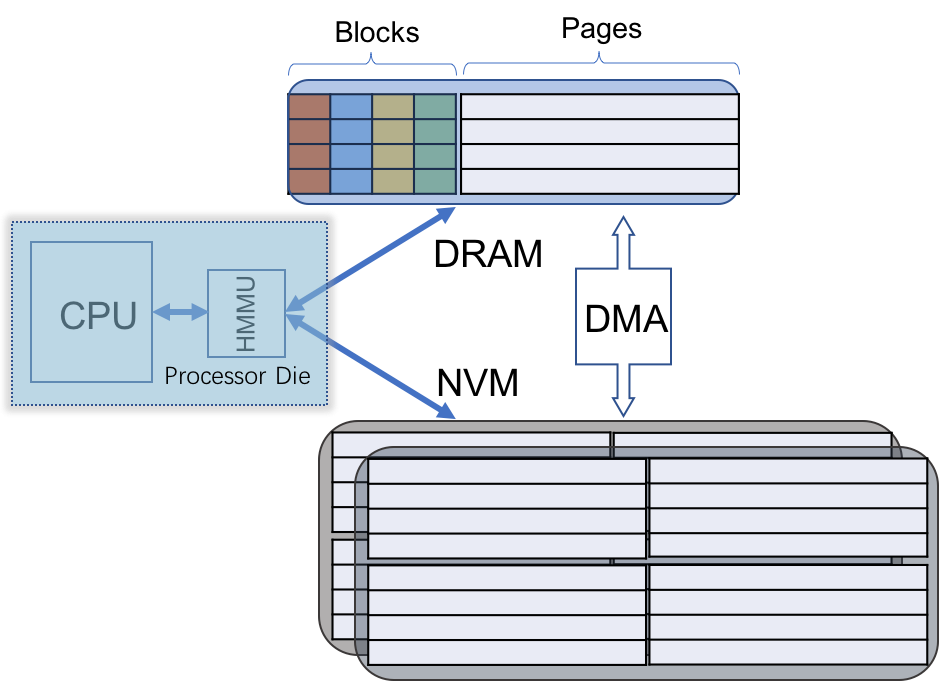}}
\caption{System Architecture Overview}
\label{fig:architecture}
\end{figure}

Figure~\ref{fig:architecture} shows the system architecture of our
proposed scheme. The data access requests are received by the Hybrid
memory management unit (HMMU), if they miss in the processor cache.
These are processed based on the built-in data placement policies, and
forwarded with address translation to either DRAM or NVM.  The HMMU
also manages the migration of data between DRAM and NVM, by
controlling the high-bandwidth DMA engine connecting the two types of
memory devices.

\subsection{Data management policy}
A key component of the proposed HMMU design is its data management
policy, \emph{i.e.} the policy by which it decides where to place and
when to move data between the different memory levels.  Traditionally,
in processor caches and elsewhere, cache blocks are managed with
64-byte lines and policies such as set-associative are used to decide
what to replace upon the insertion of new lines into a given cache
level.  While this approach yields generally good performance results
in processor caches, there are difficulties in adapting it for use in
hybrid memories.  As previously discussed in
Section~\ref{sec:dramcache}, for a hybrid memory system of 16GB
comprised of 64-byte cache-lines, the tag store overhead would be an
impractically large 2GB.  Extending the block size up to 4KB to match
the OS page size would significantly reduce the overheads of the tag
store, bringing it down to 4MB for a 16GB space.  Since the host
operating system primarily uses 4KB pages, using any larger size than
4KB for block management, however, risks moving a set of potentially
unrelated pages together in a large block, with little, if any spatial
locality between different pages in the block.  This is particularly
true because the addresses seen in the HMMU are ``physical
addresses'', thus physically colocated pages may come from completely
different applications, with no spatial relationship.\footnote{ While
  many systems do allow a subset of pages to be managed at larger
  granularities, the HMMU has no visibility to this OS-level mapping,
  thus we conservatively assume 4KB pages} As we will discuss,
however, even managing blocks on a page granularity will yield greater
than optimal page movements between ``fast'' (DRAM) and ``slow'' (NVM)
memory levels, due to the fact that only subsets of the page are ever
touched in many applications.  Thus, we will examine a hybrid scheme
in which most of the fast memory is managed on a page basis, lowering
tag overheads, while a small fraction is managed on a sub-page basis
to reduce page movement when only small portions of each page are
being used at a given time.

In terms of organization and replacement, using traditional processor
cache policies of set associativity and LRU replacement become
unwieldy for a memory system of this size.  The practical
implementation of such a set-associative cache requires either a
wide/multi-ported tag array (which becomes untenable for large SRAM
structures) or multiple cycles to retrieve and compare each way in the
set sequentially.  Prior work from the OS domain~\cite{2Q, LRUK} shows
that, with a large number of pages to choose from, set associative,
LRU replacement is not strictly necessary.  Inspired by that, we first
developed a simple counter-based page replacement policy.

\subsubsection{Counter-based Page Management}
\label{sec:counter}
\begin{figure}[!h]
  \centerline{\includegraphics[width=1.2\columnwidth]{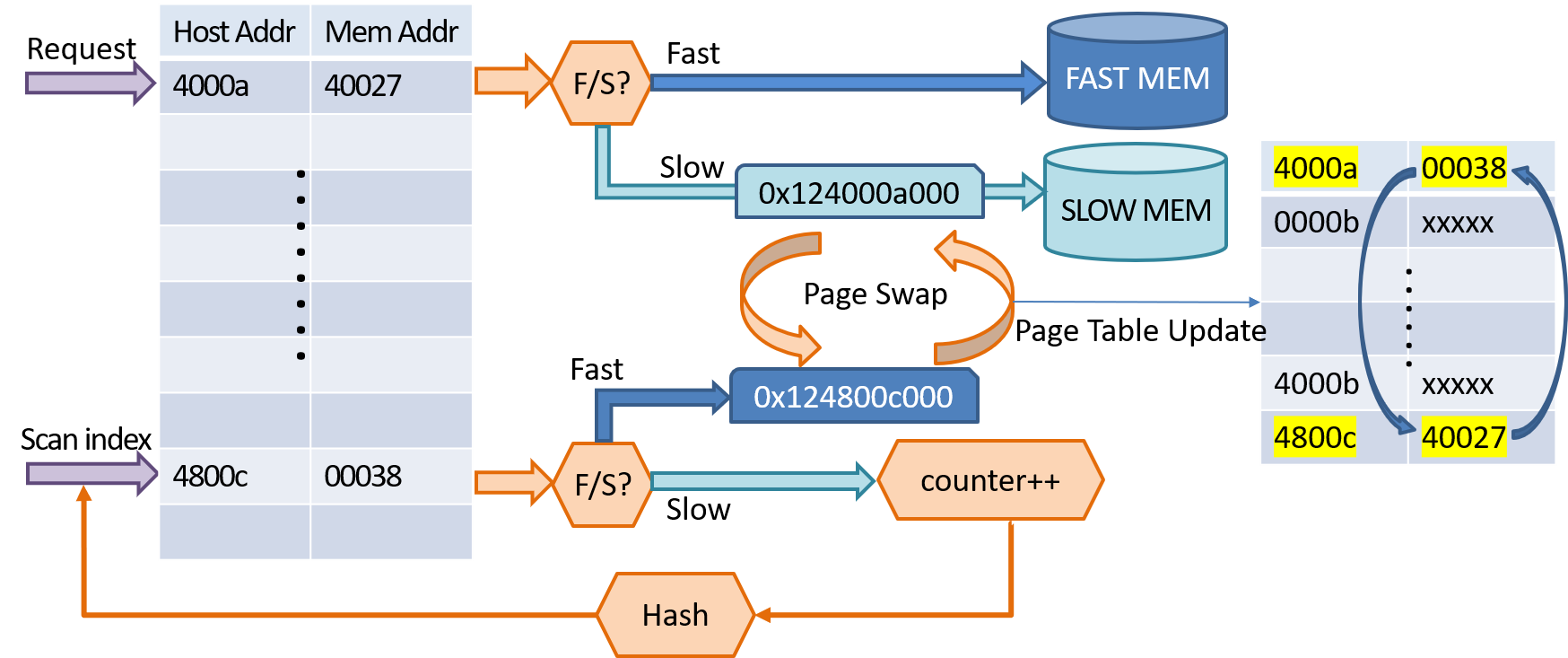}}
  \caption{Counter-based Page Movement Policy}
\label{fig:cntr-pgmv}
\end{figure}

Rather than implementing a set associative organization with the
drawbacks described above, we instead propose to implement a
secondary, page-level translation table internal to the HMMU as
illustrated in Figure~\ref{fig:cntr-pgmv}. The internal page table
provides a one-to-one remapping, associating each CPU-side
``physical'' page number in the host address space to a unique page
number in the hybrid memory address space, either in the fast or slow
memory.  Thus, any given host page can be mapped to any location in
either fast or slow memory.

While this design gives great flexibility in mapping, when a slow
memory page must be moved to fast memory (\emph{i.e.} upon a slow
memory reference we move that page to fast memory) it requires a
mechanism by which to choose the fast memory page to be replaced.
Inspired by prior work in the OS domain~\cite{2Q, LRUK}, we designed
the counter-based replacement policy for this purpose.

\paragraph{Algorithms and Design}
The counter-based replacement policy only requires one counter to keep
track of the currently selected fast memory replacement candidate
page, thus it has minimal resource overhead and can efficiently be
updated each cycle.  The chance that a recently accessed page gets
replaced is very rare because 1. the total number of pages is very
large; 2. the counter increases monotonically.  To further reduce the
possibility of evicting a recently touched page, however, we implemented
a light-weight bloom filter that tracks the last 2048 accessed pages.
Since checking against the bloom filter is parallel to normal page
scan process, and is also executed in background, it adds no extra
delay to data accesses.  Algorithm~\ref{alg:cntrpgmv} shows the
details of the algorithm.

\begin{algorithm2e}[h]
  \caption{Counter-based Page Relocation}
  \label{alg:cntrpgmv}
  \SetKwProg{Fn}{Function}{ is}{end}
  \Fn{unsigned pgtb-lookup (address)}{return page\_table[$address / page\_size$]\;}
  
  \Fn{unsigned search-free-fast-page()}{
    \While{pointed\_page  $\not\in$ fast memory {\bf or}\newline pointed\_page $\in$ bloom filter}
    {
      counter++\;
      pointed\_page \:= pagetable[Hash(counter)]\;
    }
    set candidate page as ready\;
    return pointed\_page\;
  }
  \Fn{counter-based-page-move(address)}{
    pointed\_page = pgtb-lookup (address)\;
    \eIf{pointed\_page $\in$ fast memory}{
      directly forward the request to DRAM}
    {\eIf{candidate page is available}
      {initiate to swap the content between requested page and candidate page.\;
        Call page-swap()\;}
      {
        Forward the current request to NVM\;
      }
    }
  }
  \Fn{page-swap (source\_page, target\_page)}{
    \While{page swap is not completed}{
      \If{new requests conflict with pages on flight}{Froward the requests to the corresponding device depending on the current moving progress}
      Continue the page swap\;
    }
    Update the corresponding entries in page table.\;
  }
\end{algorithm2e}

Figure \ref{fig:cntr-pgmv} illustrates a simple example of this page
movement policy.  In this example memory address space, fast memory
occupies internal page numbers 0 - 40000, while slow memory ranges
from page number 40000 and beyond.  In the figure, a request for host
address 4000a arrives at the internal remapping page table.  The
corresponding internal page address in the memory address space is
40027, which in this case is the 28th page in the slow memory.  Here
we use a policy of page movement to fast memory upon any slow memory
touch.\footnote{Note that the request is serviced immediately, directly
  from the slow memory, while the page swap happens in the
  background.} Thus, the HMMU directs the DMA engine to start
swapping data between the requested page (40027) and the destination
page in fast memory.  Here, as described in
Algorithm~\ref{alg:cntrpgmv} the fast memory pages to be replaced is
selected via the replacement counter, \emph{i.e,} page 00038 in this
example.  Once the data swapping is completed, the memory controller
updates the new memory addresses of the two swapped pages in the
internal page table.  Next, the counter searches for the next fast
memory page replacement candidate.  As the figure shows, the counter
is passed through a hash function to generate an index into the
internal page table.  If the retrieved page number turns out to be in
slow memory, the counter increments by one and the hash function
generates a new index for the next query to the page table.  Such
process loops until it finds a page in the fast memory, which becomes
the candidate destination for the page swapping.

Further details of the counter-based page management policy:
\begin{itemize}[noitemsep,nolistsep]
\item Current requests are processed at top priority under all
  circumstances.  Except for rare cases when a given write request
  conflicts with ongoing page movement, we always process the current
  request first.  As for those rare cases, since all write requests
  are treated as non-blocking, the host system shall not suspend for
  them to complete.  Therefore our design does not add overhead to the
  critical path of request processing.
\item Due to the parallel nature of hardware, we search for free pages
  in fast memory in the background, without interference to host read
  request processing.
\item Page-swap is initiated by the HMMU, however, it is executed by a
  separate DMA hardware module. Thus it does not impact other ongoing
  tasks.
\item Data coherence and consistency are maintained during page
  movements.
\end{itemize}
We carefully designed the DMA process so that it could properly handle
the new requests to the pages as they are being moved.  All read
requests and most write requests can proceed without blocking.  In
some very rare cases, the write requests are held until the current
page copy finished.
\subsubsection{Sub-page Block Management}
\label{sec:subpage}
Various applications could have widely different data access patterns:
those with high spatial locality may access a large number of adjacent
blocks of data; while others may have a larger stride between the
requested addresses.  For applications with weak or no spatial
locality, there is very limited benefit to moving the whole page of
data into fast memory, as most of the non-touched data may not be used
at all.  Based on this observation, we propose a scheme for sub-page
size block management, which manipulates the data placement and
migration in finer granularity.
\begin{algorithm2e}
\caption{Sub-page Block Management}
\label{algo:subpage}
\SetKwProg{Fn}{Function}{ is}{end}

\Fn{sub-page block management(address)}{
	pointed\_page = pgtb-lookup(address)\;
	\eIf{pointed\_page $\in$ fast memory}{
    directly forward the request to DRAM}
    {\eIf{the count of cached blocks > threshold value}{
    \eIf{candidate free page available}{initiate to swap the content between requested page and candidate page.\;
    Call page-swap()\;}{Forward the current request to NVM\;}
    }{
    initiate moving the block to cache zone
    }
}
}
\end{algorithm2e}
\begin{figure*}[!hbt]
\centerline{\includegraphics[height=0.35\textwidth]{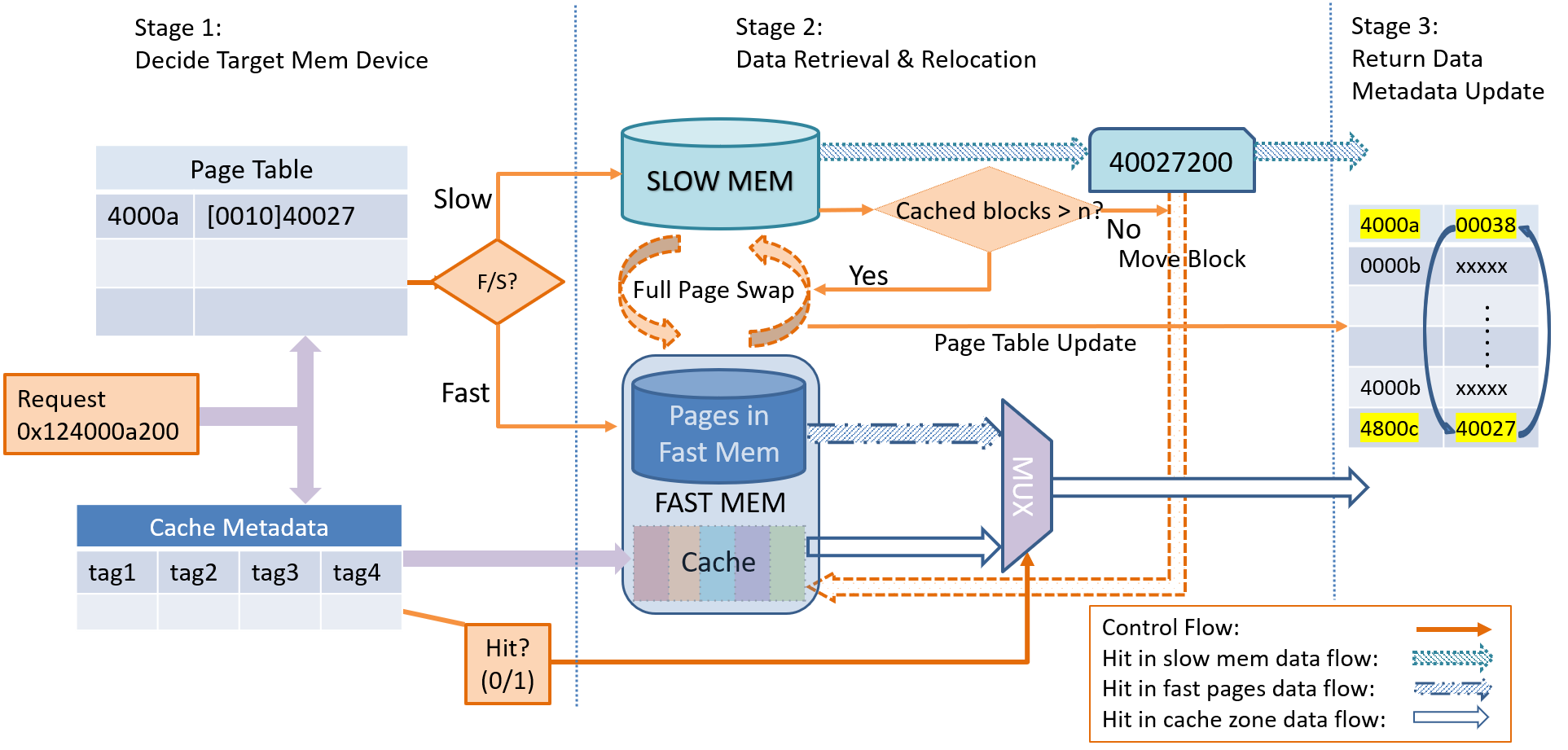}}
\caption{Sub-block Relocation Policy}
\label{fig:sub-block}
\end{figure*}
\paragraph{Data Migration Policy}
We set aside a small fraction of the fast memory and manage that area in a
cache-like fashion with sub-page sized blocks.  The basic algorithm
used in shown in Algorithm~\ref{algo:subpage}.  Upon the first
accesses to a slow memory page, instead of moving the whole page into
fast memory, we will only move the requested block of that page into
the "cache" zone in fast memory.  We then keep track of the total
number of cached blocks belonging to every page.  Only after the count
of cached blocks meets a certain threshold will we swap the whole page
to fast memory.

Figure~\ref{fig:sub-block} illustrates a simple example of the sub-page
block relocation policy: The memory controller receives a request to
host physical address 0x124000a200.  In the first cycle, both the page
table and cache metadata are checked in parallel, to decide the target
memory device.  If the data is found only in the slow memory, the
memory controller will trigger the data relocation process. The 4-bit
counter in the page table entry tells the number of sub-page blocks
that have been cached for the current page.  Comparing the counter
against the preset threshold, determines whether to start a full-page
swap or a sub-block relocation.  In the given example, the counter
value is 2, which is smaller than the threshold value of 4. Thus only
that specific block containing the requested data ( 0x40027200 to
0x4002727f) will be copied to the cache.  It is possible that the data
might be found in both slow memory and the cache at the same time.  To
enforce data consistency, we always direct the read/write request to
the copy in cache.  This dirty data will be written back to the slow
memory upon eviction.

\paragraph{Fast Memory Cache Design}
As the page size is 4KB, we choose 128 bytes as a reasonable block
size (this size also corresponds to the DRAM open page burst size, so
it sees a significant performance boost versus other block sizes).
The cache is organized as a 4-way associative cache.  
The cache uses a pseudo-LRU as the block replacement policy.  We also
enable a proactive cache recycling policy: when a block is accessed,
if its underlying page is detected to have been relocated to fast
memory, we would evict that block from the cache to save the space for
other blocks.  Thus one block of data will not occupy the capacity of two
copies in the fast memory at the same moment.
\subsubsection{Hardware Cost and Overhead}
Each page table entry takes
${\log_2 {\dfrac{\tt Memory\ Space}{\tt Page \ Size}}}$ bits to
represent the page address. In addition, we need some bits for
statistical meta-data such as the counter of misses occurring to the
page.  In our sample design, the memory space is 2GB and the page size
is 4KB, thus the hardware cost per page could be rounded to
${\log_2 {\dfrac{\tt 2GB}{\tt 4KB}}} + 5 {\tt bits} = 3{\tt bytes}$,
and the total cost is 1.5MB. The page table cost scales linearly with
memory size whereas the cost per entry only grows logarithmically.
The meta-data for each cache set is comprised of three parts; four
tags(8 bits $\times$ 4), pseudo-LRU bits (3) and dirty bits (4), which
adds to 39 bits. The total cost is
$39 {\tt bits} \times 2^{16} \approx 312{\tt KB}$ Since the cache is
read and check parallel to the access to the page table, there is no
additional timing cost for handling regular requests. The DMA provides
the non-conflict data relocation for sub-page block level as same as
that of the page relocation.

\subsubsection{Static versus Adaptive Caching Threshold}
\label{sec:adpthrs}
With both page and block migration available, a new question arises,
how to choose wisely between these two policies for optimal results.
We note that these policies have different characteristics as follows:
\begin{itemize}[noitemsep,nolistsep]
\item With page-migration, the data is exclusively placed between NVM
  and DRAM device.  Thus larger memory space is available to
  applications, and the bandwidth of both devices is available.
\item Sub-page-block migration is done in an inclusive cache fashion,
  thus avoids the additional writes to NVM when the clean data blocks
  are evicted from DRAM.
\end{itemize}
For applications with strong spatial locality, whole page migration
maximizes performance because the migration cost is only incurred
once, and the following accesses hit in the fast memory.  Alternately,
sub-page block promotion benefits applications with less spacial
locality, because it limits writes to NVM incurred by full page
migration.  We further note that application behavior may vary over
time with one policy being better in one phase and another better
during another.

We therefore include in the page translation table an 8-bit, bitmap
for tracking accesses to each sub-page block of the given page.  This
allows measurement of the utilization rate of promoted pages.  If a
large portion of blocks were revisited, then we lower the threshold to
allow more whole page migration. Alternately, if few blocks were
accessed we suppress the whole page promotion by raising the threshold
value, decreasing the rate at which full pages are
migrated.
\pgfplotsset{
    /pgfplots/ybar legend/.style={
    /pgfplots/legend image code/.code={%
       \draw[##1,/tikz/.cd,yshift=-0.25em]
        (0cm,0cm) rectangle (12pt, 1em);},
   },
}

\pgfplotstableread{
Label	slowreads	fastreads	slowwrites	fastwrites
500.perlbench	0.001101157	0.516376705	0.00008	0.482438486
510.parest	0.004104951	0.495357516	0.001382592	0.499154941
519.lbm	0.071051281	0.428234006	0.004996744	0.495717969
520.omnetpp	0.017207536	0.490861957	0.005583582	0.486346925
523.xalancbmk	0.001228754	0.497625173	0.000291998	0.500854075
531.deepsjeng	0.419667334	0.092571081	0.328708265	0.15905332
538.imagick	0.018837591	0.479471231	0.007597076	0.494094101
544.nab	0.036353339	0.46136072	0.001403656	0.500882285
557.xz	0.00527575	0.497850064	0.001835587	0.495038599
}\breakdown

\pgfplotstableread{
application	pgmv	cache	adp
500.perlbench	1	0.728698634	1.025341098
510.parest	1	1.072359193	0.947585553
519.lbm	1	0.984816135	0.77427594
520.omnetpp	1	0.140830575	0.180431568
523.xalancbmk	1	0.954352489	1.146874349
531.deepsjeng	1	1.151366501	0.641134487
538.imagick	1	0.963245939	0.897201754
544.nab	1	1.217826845	0.782726719
557.xz	1	1.027527592	1.125329377
Geomean	1	0.805992492	0.754115343
}\totalwrites

\pgfplotstableread{
threshold	runtime	writes	blockmoves	pagemoves
2	1.006885948	2.81845266	0.804516573	1.31828603
3	0.990173941	1.21211855	0.666163415	0.425627425
4	1	1	0.878191535	0.121808465
5	1.000177254	1.595548464	1.566236733	0.051588786
6	1.030198756	2.450037204	2.454505707	0.02851896

}\omnet

\pgfplotstableread{
Label   DDRbkgd NVMwrites       NVMreads        DDRreads        DDRwrites
500.perlbench.AllDRAM	0.802966823	0	0	0.110885792	0.086147385
500.perlbench.AdpComb	0.198918343	0.001519191	0.00035364	0.111307933	0.08686864
510.parest.AllDRAM	0.189231436	0	0	0.441821051	0.368947513
510.parest.AdpComb	0.066602443	0.0098196	0.002458476	0.442684964	0.372448568
519.lbm.AllDRAM	0.227013973	0	0	0.421195118	0.351790909
519.lbm.AdpComb	0.196473385	0.329126997	0.068397129	0.537131211	0.503197728
520.omnetpp.AllDRAM	0.350798573	0	0	0.359405267	0.28979616
520.omnetpp.AdpComb	0.197408397	0.050679778	0.009911628	0.389539479	0.323756828
523.xalancbmk.AllDRAM	0.440507135	0	0	0.304558739	0.254934126
523.xalancbmk.AdpComb	0.127244363	0.003080688	0.000768328	0.305270628	0.256425397
531.deepsjeng.AllDRAM	0.88399811	0	0	0.064668779	0.051333111
531.deepsjeng.AdpComb	0.283250925	0.261486781	0.059469889	0.125091496	0.159038521
538.imagick.AllDRAM	0.949274351	0	0	0.027581312	0.023144337
538.imagick.AdpComb	0.122965765	0.003436559	0.000757991	0.027861486	0.024066208
544.nab.AllDRAM	0.915277048	0	0	0.046011538	0.038711414
544.nab.AdpComb	0.094331136	0.01645967	0.003721699	0.050457004	0.045948784
557.xz.AllDRAM	0.39191154	0	0	0.333572805	0.274515655
557.xz.AdpComb	0.16430326	0.010721168	0.002413257	0.334148026	0.277318387
}\energybreakdown

\pgfplotstableread{
benchmark	pgmv	combined	adp	allDRAM
500.perlbench	0.399110689	0.400168401	0.398967746	1
510.parest	0.891711495	0.899527187	0.894014051	1
519.lbm	1.808110293	2.198493503	1.634326449	1
520.omnetpp	1.433292898	0.897765631	0.97129611	1
523.xalancbmk	0.689435006	0.693856229	0.692789406	1
531.deepsjeng	1.075926477	0.976895021	0.888337611	1
538.imagick	0.179281037	0.180051731	0.179088009	1
544.nab	0.219021698	0.249960492	0.210918294	1
557.xz	0.785082125	0.790206248	0.788904098	1
Geomean	0.651310065	0.636416319	0.601998217	1
}\energy

\section{Evaluation}
\label{sec:eval}

In this section, we present the evaluation of our proposed HMMU
design.  First, we present the experimental methodology.  Then we
discuss the performance results. Finally we analyze some of the more
interesting data points.

\subsection{Methodology}
\label{sec:method}

\subsubsection{Emulation Platform}
Evaluating the proposed system presents several unique challenges
because we aim to test the whole system stack, comprising not only the
CPU, but also the memory controller, memory devices and the
interconnections.  Further, since this project involves hybrid memory,
accurate modeling of DRAM is required.  Much of the prior work in the
processor memory domain relies upon software simulation as the primary
evaluation framework with tools such as Champsim~\cite{champsim} and
gem5~\cite{gem5}. However, detailed software simulators capable of our
goals impose huge simulation time slow-downs versus real hardware.
Furthermore, there are often questions of the degree of fidelity of
the outcome of arbitrary additions to software
simulators~\cite{7155440}.

Another alternative used by some prior work~\cite{span} is to use an
existing hardware system to emulate the proposed work.  This method
could to some extent alleviate the overlong the simulation runtime,
however, no existing system supports our proposed HMMU.

Thus, we elected to emulate the HMMU architecture on an FPGA platform.
FPGAs provide flexibility to develop and test sophisticated memory
management policies while its hardware-like nature provides near-native simulation speed. The FPGA communicates with the ARM CortexA57
CPU via a high-speed PCI Express link, and manages the two memory
modules(DRAM and NVM) directly. The DRAM and NVM memories are mapped
to the physical memory space via the PCI BAR(Base Address Register)
window.  From the perspective of the CPU, they are rendered as
available memory resource same as other regions of this unified space.

Our platform emulates various NVM access delays by adding stall cycles
to the operations executed in FPGA to access external DRAM.  The
platform is not constrained to any specific type of NVM, but rather
allows us to study and compare the behaviors across any arbitrary
combinations of hybrid memories.  In the following sections, we would
show the simulation results with different memory devices.  The
detailed system specification is listed in Table~\ref{tab:System
  Spec}.
\begin{table}[hpbt]
\footnotesize
\centering
\settowidth\tymin{\textbf{Component}}
\caption{Emulation System Specification}
\begin{tabulary}{\columnwidth}{L|L}
  \hline
  {\textbf{Component}} & Description \\
  \hline
  CPU & ARM Cortex-A57 @ 2.0GHz, 8 cores, ARM v8 architecture\\
  \hline
  L1 I-Cache & 48 KB instruction cache, 3-way set-associative\\
  \hline
  L1 D-Cache & 32 KB data cache, 2-way set-associative\\
  \hline
  L2 Cache & 1MB, 16-way associative, 64kB cache line size \\
  \hline
  Interconnection & PCI Express Gen3 (8.0 Gbps) \\
  \hline
  Memory & 128MB DDR4 + 1GB NVM \\
\hline
OS & Linux version 4.1.8 \\
\hline
\end{tabulary}
\label{tab:System Spec}
\end{table}

We measured the round trip time in FPGA cycles to access external DRAM
DIMM first, and then scaled the number of stalled cycles according to
the speed ratio between DRAM and future NVM technologies, as described in Section~\ref{sec:nvmtech}. Thus we have one DRAM DIMM running at full
speed and the other DRAM DIMM emulating the approximate speed of NVM
Memory.

\subsubsection{Workloads}

We initially considered several mobile-specific benchmark suites,
including the CoreMark~\cite{coremark} and AndEBench~\cite{AndEBench}
from EEMBC.  We found however that these suites are largely out of
date and do not accurately represent the large application footprints
found on modern mobile systems.  Also, in some cases they are only
available as closed source~\cite{AndEBench} and thus are unusable in
our infrastructure.  Instead, we use applications from the recently
released SPEC CPU 2017 benchmark suite~\cite{SPEC_Official}.  To
emulate memory intensive workloads for future mobile space, we selected
only those SPEC CPU 2017 benchmarks which require a larger working set
than the fast memory size in our system.  The details of tested
benchmarks are listed in Table~\ref{tab:Benchmarks}.

To ensure that application memory was allocated to the HMMU's memory,
the default Linux malloc functions are replaced with a customized
jemalloc~\cite{jemalloc}.  Thus the HMMU memory access was transparent
to the CPU and cache, and no benchmark changes were needed.

\begin{table}[hpbt]
\footnotesize
\centering
\settowidth\tymin{\textbf{Memory Footprint}}
\caption{Tested Workloads\cite{SPEC_Official}} \label{tab:Benchmarks}

\begin{tabulary}{\columnwidth}{L|L|L}
\hline
{\textbf{Benchmark}} & Description & Memory footprint \\
\hline
\multicolumn {3}{c}{Integer Application} \\
\hline
500.perlbench	& Perl interpreter & 202MB \\
\hline
520.omnetpp	& Discrete Event simulation - computer network	& 241MB \\
\hline
523.xalancbmk	& XML to HTML conversion via XSLT & 481MB \\
\hline
531.deepsjeng	& Artificial Intelligence: alpha-beta tree search (Chess) & 700MB \\
\hline
557.xz & General data compression & 727MB \\
\hline
\multicolumn {3}{c}{Float Point Application} \\
\hline
510.parest & Biomedical imaging: optical tomography with finite elements & 413MB \\
\hline
519.lbm & Fluid dynamics & 410MB \\
\hline
538.imagick & Image Manipulation & 287MB \\
\hline
544.nab & Molecular Dynamics & 147MB \\
\end{tabulary}
\end{table}

\subsubsection{Designs Under Test}
Here we test the following data management policies developed for use
with our HMMU:
\begin{itemize}[noitemsep,nolistsep]
\item \textbf{Static:} A baseline policy in which host requested pages
  are randomly assigned to fast and slow memory.  This serves as a
  nominal, worst-case, memory performance.
\item \textbf{PageMove:} The whole 128MB DRAM is manged on the
  granularity of 4k pages.  When a memory request is missed in fast
  memory, the DMA engine will trigger a page relocation from slow
  memory to fast memory, as described in Section~\ref{sec:counter}.
\item \textbf{StatComb:} Here 16MB out of the 128MB DRAM is reserved
  for sub-page block relocation, managed in the cache-like fashion, as
  described in Section~\ref{sec:subpage}.  The remainder of the DRAM
  is managed on a full page basis. An empirically derived static
  threshold of 4 blocks touched is used to determine when a full page
  should be moved to the page portion of DRAM.
\item \textbf{AdpComb:} Same as StatComb, except that, as described in
  Section~\ref{sec:adpthrs}, an adaptive threshold is used to
  determine when the full page should be moved.
\item \textbf{AllDRAM:} Here we implement a baseline policy in which
  there is sufficient fast memory to serve all pages in the system and
  no page movement is required.  This serves as a nominal, best-case
  but impractical memory performance design.
\end{itemize}

\subsection{Results}
\label{energy-analysis}
\subsubsection{Energy Saving}
Emerging NVM consumes minimal standby power, which could help save
energy consumption on mobile computation.  We evaluated and compared
the energy spent in running SPEC 2017 benchmarks between the full DRAM
configuration and our policies.  We referred to Micron DDR4 technical
spec~\cite{micron-ddr4} for DRAM and recent work on
3DxPoint~\cite{nvm-power} for NVM device power consumption,
respectively (Table~\ref{dram-3dp-power}).
\begin{table}[!hbt]
\centering
\caption{Power Consumption of DDR4 and 3D-XPoint}
\begin{tabular}{c|c|c}
\hline
Technology& DDR4 & 3Dxpoint\\
\hline
Read Latency & 50ns & 100ns \\
Write Latency & 50ns & 300ns \\
Read Energy & 4.2nJ & 1.28nJ \\
Write Energy & 3.5nJ & 8.7nJ \\
Background Power & 30mW/GB & $\sim 0$ \\
\hline 
\end{tabular}
\label{dram-3dp-power}
\end{table}

\begin {figure}[!hbt]
  \centering
  \includegraphics{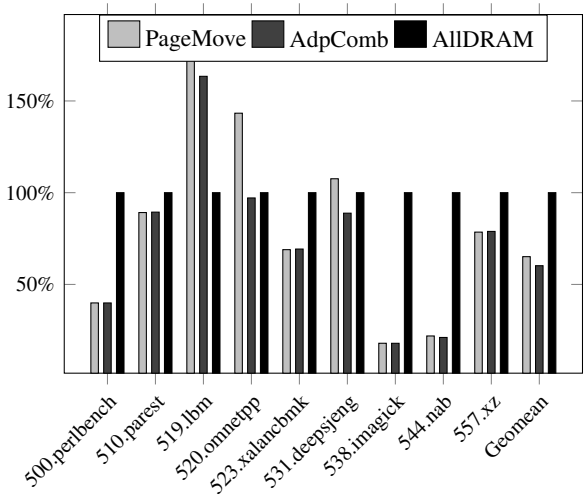}
  \caption{Energy Consumption Comparison} 
  \label{fig:energy}
\end{figure}

We normalize the energy consumption of our policies to that of the
AllDRAM configuration and present them in the figure~\ref{fig:energy}.
In the figure we see that all three techniques save a substantial
amount of energy.  On average the AdpComb adaptive policy only
consumes 60.2\% energy as compared to the AllDRAM configuration, while
the PageMove and StatComb policies are at 65.1\% and 63.6\%,
respectively.  That said, several benchmarks see energy consumption
increases under the PageMove policy, while StatComb, sees a
significant regression in energy consumption for 519.lbm. AdpComb,
while also seeing increased energy consumption under 519.lbm, shows
better energy consumption than the other two policies for nearly all
cases.

\begin {figure}[!hbt]
  \centering
  \includegraphics[width=0.99\linewidth]{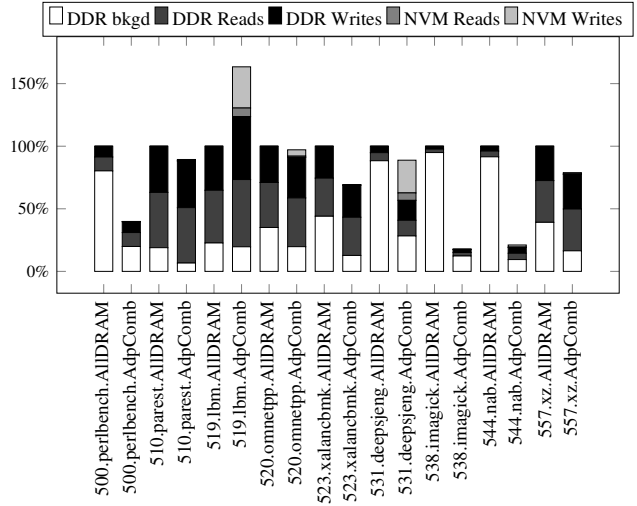}
  \caption{Energy Consumption Breakdown}
  \label{fig:energybreakdown}
\end{figure}
Further investigating the distribution of energy consumption, we track
the DRAM background power, number of DRAM read/writes and NVM
read/writes.  We present the comparison between AdpComb and AllDRAM in
Figure ~\ref{fig:energybreakdown}.  Since 7/8 of the memory was replaced
with NVM, the standby power shrinks significantly.  Although write
operations to NVM dissipate more energy than DRAM, the AdpComb policy
avoids most of this increase by absorbing many writes in DRAM.  Our
policies saw the greatest energy efficiency improvement with
applications imagick and nab, which spent 17.9\% and 21.1\% energy
compared to full DRAM.  We find that these two applications have high
processor cache hit rates and spent most time in computation.  Thus
they have few references to the memory, and the largest portion of
energy was spent on DRAM background power.  Thus AptComb policy's
advantage of having much lower DRAM static power is best exploited.
Our policies did pretty well with all benchmark applications except
lbm, which spent 63\% more energy.  This application incurred a massive
number of cache block writebacks to NVM.  We investigated the case and
found lbm has the highest percentage of store instructions among all
benchmark applications~\cite{spec2017-workload}. This creates many
dirty blocks, and thus writebacks are expected when blocks are later
evicted.  The amount of writes is also amplified by the writebacks of
cache blocks.
\label{sec:speedup}
\subsubsection{Runtime Performance}
\begin{figure*}
  \centering
  \includegraphics{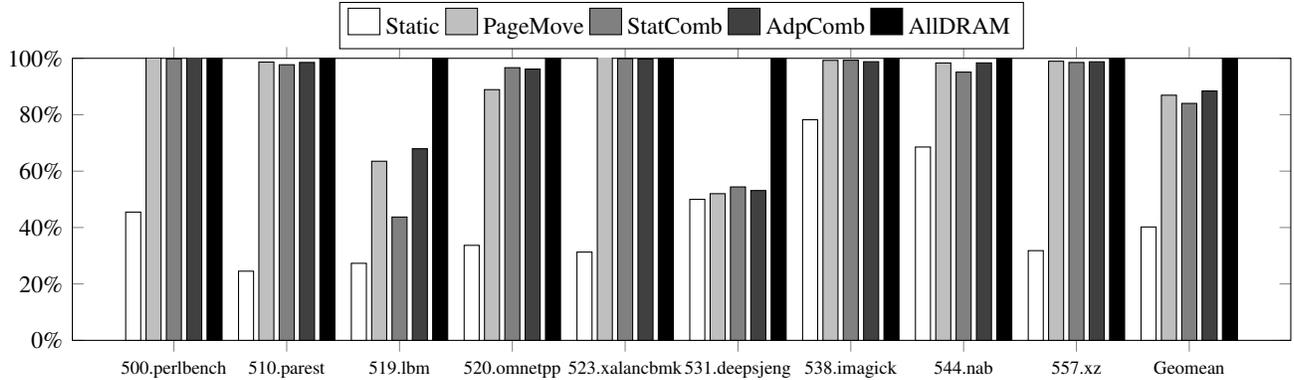}
\caption{SPEC 2017 Performance Speedup}
\label{fig:speedup}
\end{figure*}

Figure~\ref{fig:speedup} shows the speedup attained by the different
designs under test for the various benchmarks in the SPEC CPU 2017
benchmark suite.  Here all the results are normalized to the runtime of
the ideal, AllDRAM, DRAM configuration.  We see that the average
performance of AdpComb is 88.4\%, while the random static allocation
``Static'' only yields 40\% of the full DRAM performance.  Thus, the
adaptive policy achieves more than 2x performance benefit versus the
worst-case, static allocation policy under the same memory resource.
Generally the AdpComb policy outperforms the other two policies we
propose, though interestingly, for many benchmarks, including
perlbench, parest, xalancbmk, xz, imagick, and nab, PageMove comes
within 5\% of the performance of AllDRAM.

\subsection{Analysis and Discussion}
The adaptive AdpComb policy successfully reduces energy by 40\%, with
a modest 12\% loss of the performance versus an unrealistic and
unscalable AllDRAM design.  AdpComb attempts to make the optimal
choice between the PageMove and the StatComb block migration policy.
In the remainder of this text, we will further analyze the experiment
results.

\subsubsection{PageMove Policy Performance}
\begin {figure}[hbt]
  \centering
  \includegraphics[width=\columnwidth]{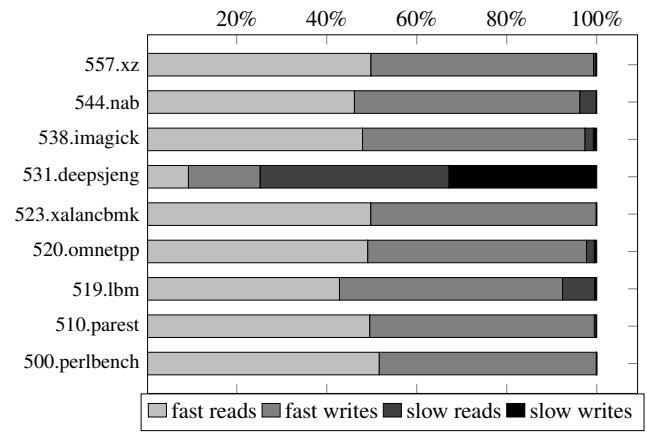}
  \caption{Memory Accesses Breakdown of PageMove Policy} 
  \label{fig:memory_hits}
\end{figure}
The PageMove policy has similar average runtime performance (86.9\%)
to the adaptive AdpComb policy (88.4\%).  Figure~\ref{fig:memory_hits}
shows the breakdown of memory requests that hit in the fast pages and
slow pages respectively.  When compared to the speedup in
Figure~\ref{fig:speedup}, we see the benchmarks which PageMove policy
works best have most of their memory requests hitting in the fast
pages, while the hit rate in slow pages become negligible. This
provides a large performance boost considering that the system's slow
memory is 8x slower than the fast memory.

In the figure, the StatComb policy has an overall speedup of 84\%
against the AllDRAM configuration. The difference is mainly
contributed by 519.lbm and 544.nab.  As we will show, however,
StatComb does still provide significant benefits in terms of total
writes to NVM.

The PageMove policy performs worst on the benchmark 531.deepsjeng, with a slowdown of 52\% versus AllDRAM.  We divided the number of hits in
fast memory by the occurrences of page relocation, and found that
deepsjeng has the lowest rate (0.03) across all the benchmark
applications (Geomean is 3.96).  This suggests that when a page is
relocated from slow memory to fast memory, the remainder of that page
is often not extensively utilized.  Further, we also see an
exceptionally high ratio of blocks moved to cache versus page
relocation.  The geometric mean of all benchmarks is 10.5 while
deepsjeng marks 397.  This is a sign that in most cases, the page is
only visited for one or two lines, and never accumulates enough cached
blocks to begin a whole page relocation.  To sum up, deepsjeng has a
sparse and wide-range memory access pattern, which is quite difficult
to prefetch effective data or improve performance.

519.lbm presents another interesting case, since its performance is
also poor.  Similar to deepsjeng, the hit rate in fast memory is low
in contrast to the number of page relocations.  However, a key
difference is that over 60\% of the cached blocks were evicted after
its underlying pages relocated to fast memory.  This indicates that
lbm walks through many blocks of the same page and triggers the whole
page relocation quickly.  On that account, we deduce that this
benchmark will benefit from a configuration with more fast pages and a
smaller cache zone.  We reran this benchmark with a cache size of 8MB
and the threshold value of 1, and found a supportive result of 8\%
performance gain on top of the default threshold value of 4.

\subsubsection{Writes Reduction and NVM lifetime Saving}
\begin {figure}[!hbt]
\includegraphics{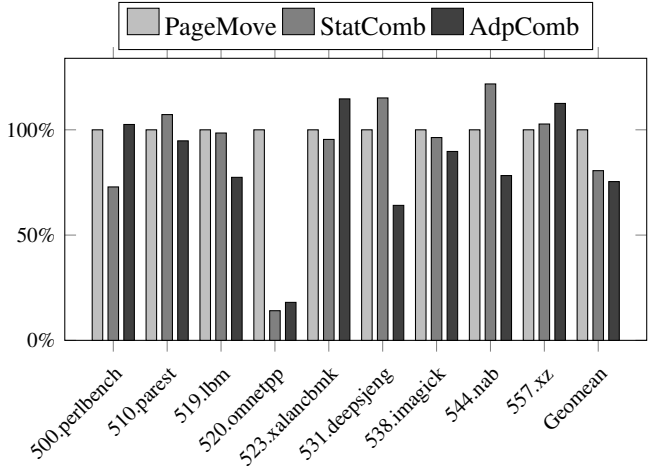}
\caption{Writes to NVM}
\label{fig:writes_nvm}
\end{figure}
Unlike the traditional DRAM, emerging NVM technologies have different
characteristics for reads and writes.  Write operations dissipates
more than 8x the energy of reads~\cite{NVM1}.  Moreover, NVM
technologies often have limited write endurance, i.e, the maximum
cycles of writes before they wear out.  Hence, if we could reduce the
amount of writes, we could greatly save energy consumption and extend
the lifetime of NVM device.  Figure~\ref{fig:writes_nvm} shows the
percentage of writes to slow memory for both techniques, normalized
against the number of writes seen in the PageMove policy. Please note
that we measure not only the direct writes from the host but also the
writes induced by page movements and sub-page block writebacks to slow
memory.  In the figure we see that our combined policy has an average
of 20\% fewer writes than the PageMove policy.  While several
benchmarks benefit from the sub-page block cache, this advantage is
strongest with omnetpp, with a drop of 86\%.  The detailed analysis of
this particular benchmark is presented in the next section.
\subsubsection{Sensitivity to Threshold}

\begin {figure}[!hbt]
  \includegraphics{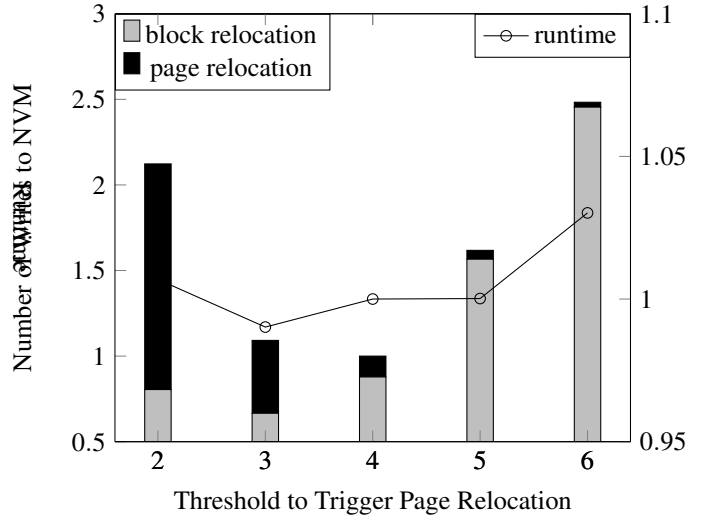}
  \caption{Omnetpp Performance Analysis}
\label{fig:omnetpp}
\end{figure}

The extraordinary reduction of writes for omnetpp is intriguing.  We
reran the tests with different StatComb static page relocation
thresholds and examined the changes in run time and total numbers of
writes to NVM.  In Figure~\ref{fig:omnetpp}, we normalized
all numbers to the value for a threshold of 4, the threshold used in
StatComb.  The runtime varied according to the same trend as the
number of writes, and the threshold value of 4 turned out to be the
overall sweet spot.  Both metrics started to deteriorate rapidly when
the threshold value shifted.  Then we measured the number of writes to
NVM incurred by page relocation and block relocation, respectively.
The results represented by stacked bars, reveals the reason
why threshold of 4 is the best choice: More pages were relocated when
the threshold was lowered.  On the other hand, the amount of block
migration grew rapidly as the threshold increased.  The trade-offs
reached perfect balance at the value of 4, which had a slightly more
page moves than that of value 5, yet significantly fewer block
migrations.
\subsubsection{Adaptive Policy}
The analysis above showed that the whole page promotion policy favors
certain benchmark applications, in which most blocks were revisited on
the promoted pages.  Meanwhile other applications benefit from
sub-page block promotions as only a subset of blocks were re-utilized.
If we could always choose the correct policy for each application, then
we could expect the optimal results for overall performance.  These
results reinforce the reasoning behind our AdpComb policy's adaptive
threshold, wherein for applications where pages are mostly utilized
full page movement is completed quickly, while for applications where
accesses are sparse, page movement is postponed till most of the page
has been touched once.
\section{Conclusions}
\label{sec:conc}
A wide spectrum of non-volatile memory (NVM) technologies are
emerging, including phase-change memories (PCM), memristor, and 3D
XPoint.  These technologies look particularly appealing for inclusion
in the mobile computing memory hierarchy.  While NVM provides higher
capacity and less static power consumption, than traditional DRAM, its
access latency and write costs remain problematic.  Integration of
these new memory technologies in the mobile memory hierarchy requires
a fundamental rearchitecting of traditional system designs.  Here we
presented a hardware-accelerated memory manager that addresses both
types of memory in a flat space address space.  We also designed a set
of data placement and data migration policies within this memory
manager, such that we may exploit the advantages of each memory
technology. While the page move policy provided good performance,
adding a sub-page-block caching policy helps to reduce writes to NVM
and save energy.  On top of these two fundamental policies, we built
an adaptive policy that intelligently chooses between them, according
to the various phases of the running application.  Experimental
results show that our adaptive policy can significantly reduce 
power consumption by almost 40\%.  With only a small fraction of the
system memory implemented in DRAM, the overall system performance
comes within 12\% of the full DRAM configuration, which is more than
2X the performance of random allocation of NVM and DRAM. By reducing
the number of writes to NVM, our policy also helps to extend device
lifetime.

\bibliographystyle{plain}
\bibliography{ref}

\begin{thebibliography}{10}

\bibitem{gem5}
Nathan Binkert, Bradford Beckmann, Gabriel Black, Steven~K. Reinhardt, Ali
  Saidi, Arkaprava Basu, Joel Hestness, Derek~R. Hower, Tushar Krishna, Somayeh
  Sardashti, Rathijit Sen, Korey Sewell, Muhammad Shoaib, Nilay Vaish, Mark~D.
  Hill, and David~A. Wood.
\newblock The gem5 simulator.
\newblock {\em SIGARCH Comput. Archit. News}, 39(2):1--7, August 2011.

\bibitem{champsim}
ChampSim.
\newblock Champsim, 2016.
\newblock \url{https://github.com/ChampSim/ChampSim}.

\bibitem{NVM1}
An~Chen.
\newblock A review of emerging non-volatile memory (nvm) technologies and
  applications.
\newblock {\em Solid-State Electronics}, 125:25--38, 2016.

\bibitem{techinsights}
Jeongdong Choe.
\newblock Intel 3d xpoint memory die removed from intel optane pcm, 2017.
\newblock
  \url{https://www.techinsights.com/blog/intel-3d-xpoint-memory-die-removed-intel-optanetm-pcm-phase-change-memory}.

\bibitem{CAMEO}
C.~C. Chou, A.~Jaleel, and M.~K. Qureshi.
\newblock Cameo: A two-level memory organization with capacity of main memory
  and flexibility of hardware-managed cache.
\newblock In {\em 2014 47th Annual IEEE/ACM International Symposium on
  Microarchitecture}, pages 1--12, Dec 2014.

\bibitem{AndEBench}
EEMBC.
\newblock An eembc benchmark for android devices, 2015.
\newblock \url{http://www.eembc.org/andebench}.

\bibitem{memristor}
K.~Eshraghian, Kyoung-Rok Cho, O.~Kavehei, Soon-Ku Kang, D.~Abbott, and
  Sung-Mo~Steve Kang.
\newblock Memristor mos content addressable memory (mcam): Hybrid architecture
  for future high performance search engines.
\newblock {\em Very Large Scale Integration (VLSI) Systems, IEEE Transactions
  on}, 19(8):1407--1417, Aug 2011.

\bibitem{jemalloc}
Jason Evans.
\newblock Jemalloc, 2016.
\newblock \url{http://jemalloc.net/}.

\bibitem{span}
Viacheslav Fedorov, Jinchun Kim, Mian Qin, Paul~V. Gratz, and A.~L.~Narasimha
  Reddy.
\newblock Speculative paging for future nvm storage.
\newblock In {\em Proceedings of the International Symposium on Memory
  Systems}, MEMSYS '17, pages 399--410, New York, NY, USA, 2017. ACM.

\bibitem{Hassan:2015}
Ahmad Hassan, Hans Vandierendonck, and Dimitrios~S. Nikolopoulos.
\newblock Software-managed energy-efficient hybrid dram/nvm main memory.
\newblock In {\em Proceedings of the 12th ACM International Conference on
  Computing Frontiers}, CF '15, pages 23:1--23:8, New York, NY, USA, 2015. ACM.

\bibitem{tag-reduction}
Cheng-Chieh Huang and Vijay Nagarajan.
\newblock Atcache: Reducing dram cache latency via a small sram tag cache.
\newblock In {\em Proceedings of the 23rd International Conference on Parallel
  Architectures and Compilation}, PACT '14, pages 51--60, New York, NY, USA,
  2014. ACM.

\bibitem{intel-nvmessd}
{INTEL CORPORATION}.
\newblock Intel 750, 2015.
\newblock
  \url{https://ark.intel.com/products/86740/Intel-SSD-750-Series-400GB-12-Height-PCIe-3_0-20nm-MLC}.

\bibitem{intel-3dxpoint}
{INTEL CORPORATION}.
\newblock Intel optane technology, 2016.
\newblock
  \url{https://www.intel.com/content/www/us/en/architecture-and-technology/intel-optane-technology.html}.

\bibitem{Unison-cache}
D.~Jevdjic, G.~H. Loh, C.~Kaynak, and B.~Falsafi.
\newblock Unison cache: A scalable and effective die-stacked dram cache.
\newblock In {\em 2014 47th Annual IEEE/ACM International Symposium on
  Microarchitecture}, pages 25--37, Dec 2014.

\bibitem{2Q}
Theodore Johnson and Dennis Shasha.
\newblock 2q: A low overhead high performance buffer management replacement
  algorithm.
\newblock In {\em VLDB}, 1994.

\bibitem{Strata}
Youngjin Kwon, Henrique Fingler, Tyler Hunt, Simon Peter, Emmett Witchel, and
  Thomas Anderson.
\newblock Strata: A cross media file system.
\newblock In {\em Proceedings of the 26th Symposium on Operating Systems
  Principles}, SOSP '17, pages 460--477, New York, NY, USA, 2017. ACM.

\bibitem{nvm-power}
Benjamin~C. Lee, Engin Ipek, Onur Mutlu, and Doug Burger.
\newblock Architecting phase change memory as a scalable dram alternative.
\newblock In {\em Proceedings of the 36th Annual International Symposium on
  Computer Architecture}, ISCA '09, pages 2--13, New York, NY, USA, 2009. ACM.

\bibitem{Tagless}
Y.~Lee, J.~Kim, H.~Jang, H.~Yang, J.~Kim, J.~Jeong, and J.~W. Lee.
\newblock A fully associative, tagless dram cache.
\newblock In {\em 2015 ACM/IEEE 42nd Annual International Symposium on Computer
  Architecture (ISCA)}, pages 211--222, June 2015.

\bibitem{spec2017-workload}
A.~Limaye and T.~Adegbija.
\newblock A workload characterization of the spec cpu2017 benchmark suite.
\newblock In {\em 2018 IEEE International Symposium on Performance Analysis of
  Systems and Software (ISPASS)}, pages 149--158, 2018.

\bibitem{Liu:2017}
Haikun Liu, Yujie Chen, Xiaofei Liao, Hai Jin, Bingsheng He, Long Zheng, and
  Rentong Guo.
\newblock Hardware/software cooperative caching for hybrid dram/nvm memory
  architectures.
\newblock In {\em Proceedings of the International Conference on
  Supercomputing}, ICS '17, pages 26:1--26:10, New York, NY, USA, 2017. ACM.

\bibitem{DRAM-cache1}
N.~Madan, L.~Zhao, N.~Muralimanohar, A.~Udipi, R.~Balasubramonian, R.~Iyer,
  S.~Makineni, and D.~Newell.
\newblock Optimizing communication and capacity in a 3d stacked reconfigurable
  cache hierarchy.
\newblock In {\em 2009 IEEE 15th International Symposium on High Performance
  Computer Architecture}, pages 262--274, Feb 2009.

\bibitem{Meza}
J.~Meza, J.~Chang, H.~Yoon, O.~Mutlu, and P.~Ranganathan.
\newblock Enabling efficient and scalable hybrid memories using
  fine-granularity dram cache management.
\newblock {\em IEEE Computer Architecture Letters}, 11(2):61--64, 2012.

\bibitem{Dram-cache2}
J.~Meza, J.~Chang, H.~Yoon, O.~Mutlu, and P.~Ranganathan.
\newblock Enabling efficient and scalable hybrid memories using
  fine-granularity dram cache management.
\newblock {\em IEEE Computer Architecture Letters}, 11(2):61--64, 2012.

\bibitem{micron-ddr4}
Inc. Micron~Technology.
\newblock Calculating memory power for ddr4 sdram.
\newblock Technical report, 2017.

\bibitem{NVM2}
S.~Mittal and J.~S. Vetter.
\newblock A survey of software techniques for using non-volatile memories for
  storage and main memory systems.
\newblock {\em IEEE Transactions on Parallel and Distributed Systems},
  27(5):1537--1550, 2016.

\bibitem{7155440}
T.~Nowatzki, J.~Menon, C.~Ho, and K.~Sankaralingam.
\newblock Architectural simulators considered harmful.
\newblock {\em IEEE Micro}, 35(6):4--12, Nov 2015.

\bibitem{LRUK}
Elizabeth~J. O'Neil, Patrick~E. O'Neil, and Gerhard Weikum.
\newblock The lru-k page replacement algorithm for database disk buffering.
\newblock In {\em Proceedings of the 1993 ACM SIGMOD International Conference
  on Management of Data}, SIGMOD '93, pages 297--306, New York, NY, USA, 1993.
  ACM.

\bibitem{Alloy-cache}
M.~K. Qureshi and G.~H. Loh.
\newblock Fundamental latency trade-off in architecting dram caches:
  Outperforming impractical sram-tags with a simple and practical design.
\newblock In {\em 2012 45th Annual IEEE/ACM International Symposium on
  Microarchitecture}, pages 235--246, Dec 2012.

\bibitem{Qureshi:2009}
Moinuddin~K. Qureshi, Vijayalakshmi Srinivasan, and Jude~A. Rivers.
\newblock Scalable high performance main memory system using phase-change
  memory technology.
\newblock In {\em Proceedings of the 36th Annual International Symposium on
  Computer Architecture}, ISCA '09, pages 24--33, New York, NY, USA, 2009. ACM.

\bibitem{PCM}
S.~Raoux, G.~W. Burr, M.~J. Breitwisch, C.~T. Rettner, Y.~. Chen, R.~M. Shelby,
  M.~Salinga, D.~Krebs, S.~. Chen, H.~. Lung, and C.~H. Lam.
\newblock Phase-change random access memory: A scalable technology.
\newblock {\em IBM Journal of Research and Development}, 52(4.5):465--479, July
  2008.

\bibitem{ITRS2015}
INTERNATIONAL TECHNOLOGY ROADMAP~FOR SEMICONDUCTORS.
\newblock Moremoore, 2015.
\newblock
  \url{https://www.semiconductors.org/resources/2015-international-technology-roadmap-for-semiconductors-itrs/}.

\bibitem{coremark}
Markus~Levy Shay Gal-On.
\newblock Exploring coremark - a benchmark maximizing simplicity and efficacy,
  2012.
\newblock \url{https://www.eembc.org/techlit/articles/coremark-whitepaper.pdf}.

\bibitem{nvm_price}
Anton Shilov.
\newblock Pricing of intel's optane dc persistent memory modules, 2019.
\newblock
  \url{https://www.anandtech.com/show/14180/pricing-of-intels-optane-dc-persistent-memory-modules-leaks}.

\bibitem{J.Sim}
J.~Sim, A.~R. Alameldeen, Z.~Chishti, C.~Wilkerson, and H.~Kim.
\newblock Transparent hardware management of stacked dram as part of memory.
\newblock In {\em 2014 47th Annual IEEE/ACM International Symposium on
  Microarchitecture}, pages 13--24, Dec 2014.

\bibitem{SPEC_Official}
SPEC.
\newblock {SPEC CPU2017 Documentation}, 2017.
\newblock \url{https://www.spec.org/cpu2017/Docs/}.

\bibitem{Wang:2014}
Z.~Wang, Z.~Gu, and Z.~Shao.
\newblock Optimizated allocation of data variables to pcm/dram-based hybrid
  main memory for real-time embedded systems.
\newblock {\em IEEE Embedded Systems Letters}, 6(3):61--64, Sept 2014.

\bibitem{Wu:2009}
X.~Wu and A.~L.~N. Reddy.
\newblock Managing storage space in a flash and disk hybrid storage system.
\newblock In {\em 2009 IEEE International Symposium on Modeling, Analysis
  Simulation of Computer and Telecommunication Systems}, pages 1--4, Sept 2009.

\bibitem{yang:2012}
J.~Joshua Yang, Dmitri~B. Strukov, and Duncan~R. Stewart.
\newblock Memristive devices for computing.
\newblock {\em Nature Nanotechnology}, Dec 2012.

\bibitem{FlashShare}
Jie Zhang, Miryeong Kwon, Donghyun Gouk, Sungjoon Koh, Changlim Lee, Mohammad
  Alian, Myoungjun Chun, Mahmut~Taylan Kandemir, Nam~Sung Kim, Jihong Kim, and
  Myoungsoo Jung.
\newblock Flashshare: Punching through server storage stack from kernel to
  firmware for ultra-low latency ssds.
\newblock In {\em 13th {USENIX} Symposium on Operating Systems Design and
  Implementation ({OSDI} 18)}, pages 477--492, Carlsbad, CA, 2018. {USENIX}
  Association.

\end{thebibliography}

\end{document}